\documentclass[showpacs,superscriptaddress, preprintnumbers,amsmath,amssymb]{revtex4}

\usepackage{graphicx}
\usepackage{dcolumn}
\usepackage{bm}
\include{epsf}

\begin{document}


\title{
iNFORMATiON SOCiETY: \\ MODELiNG A COMPLEX
SYSTEM WiTH SCARCE DATA}


\author{Noemi L. Olivera}\email{nolivera@jursoc.unlp.edu.ar}
\affiliation{ Grupo de Estudio de la Complejidad en la Sociedad de la informaci\' on (GECSi)
Facultad de Ciencias Jur'dicas y Sociales, Universidad Nacional de La Plata,
Argentina}
\author{Araceli N. Proto }\email{aproto@fi.uba.ar}
\affiliation{ Comisi\'on de investigaciones Cient'ficas de la Provincia de Buenos Aires;
Laboratorio de Sistemas Complejos, Facultad de ingenier'a, Universidad de Buenos
Aires, Argentina} 
\author{Marcel Ausloos }\email{marcel.ausloos@ulg.ac.be}
\affiliation{Group for Research and Applications of Physics in Economy and Sociology (GRAPES)\\SUPRATECS, ULG,   Sart-Tilman, B-4000 Li\`ege, Euroland
}

\date{today}

\begin{abstract}
Considering electronic implications in the Information Society (IS) as a complex system,  complexity
science tools are used to describe the processes that are seen to be taking place. The sometimes
troublesome relationship between the information and communication new
technologies and e-society gives rise to different problems, some of them being
unexpected. Probably, the Digital Divide (DD) and the internet Governance (IG) are
among the most conflctiive ones of internationally based e-Affairs. Admitting that  solutions should be found for
these problems, certain international policies are required. in this context, data  gathering and subsequent analysis, as well as
the construction of adequate physical models are extremely important in order to
imagine different future scenarios and suggest some subsequent control. in the main text, mathematical
modelization helps for visualizing how policies could e.g. influence the individual and collective behavior in an empirical
  social agent system. in order to show how this purpose could be achieved, two  approaches, (i) the Ising model and (ii) a generalized Lotka-Volterra model  are used for DD and IG considerations
respectively. it can be concluded that the social modelization of the e-Information Society as a
complex system provides insights about how DD can be reduced and
how the a large number of weak members of the IS could influence the outcomes
of the  IG.
\end{abstract}



\maketitle

\section{introduction}\label{intro}

The intensive usage of the information and communication technologies (ICT) in
daily life has given rise to the Information Society (IS) concept, as well as to different
problems, some of them unexpected, for which it is clear that stable solutions should
be found as soon as possible. Therefore certain policies are rapidly required. Neither
their plan nor their regulation should be left to theoretical considerations only, but
could be surely much focussed if some primary testing could be made before their
implementation.
Due to the global ICT interconnections, information is transmitted almost
instantaneously. But different societies and cultures have different reactions to the
information received and different times to absorb it. In order to reach a Òpeople centered,
inclusive and development-oriented Information SocietyÓ the World Summit
on the Information Society proclaims (WSIS-03/GENEVA/DOC/4-E, 12 December
2003), policies to be applied should be carefully analyzed as they need to fill up at
least two different issues: (a) to be acceptable for the national society, and (b) to be
coherent with the external conditions.

The present study concerning the electronic aspect of the IS, here below so called e-Information Society (eIS), is aimed at contributing to the
definition of the most adequate policies that are required to achieve a harmonious IS.
This contribution is devoted to show how social, legal and ethical aspects of the IS
can and should be analyzed using the modern concepts of complex systems, as the eIS has most
of those systems properties, in the sense that societies and economies are interrelated
in a nonlinear way and often self-organizing within some general constraints. For this
reason the socioeconomic problems which are spread globally need solutions which
should be adequate to each society. This is not a minor task: modelization to forecast
future scenarios is necessary, as it is expected to give at least a preliminary idea about the
consequences of proposed 'solutions', especially since the speed at which interactions
takes place, on the e-markets or more generally through e-informations, in the presently globalized context is quasi infinite. Moreover due to the new
ICT, the context operates world wide on line, thus  not only at the intranational but also at the international level.

Two relevant problems regarding the eIS are treated in this contribution: (a) the Digital
Divide (DD) and (b) internet Governance (IG), both to be further defined and briefly
commented upon in Section 2. Up to now, these two problems have been mostly expressed
only through the perception  of policy makers and intuition of social scientists, - and there is little
data to be analyzed.

\begin{itemize} \item
On one hand, measurements on DD demand, to begin with, a  strict  definition of
the concept, and afterwards, both careful design of polls and clear specifications about
what is to be measured. Unfortunately, the existing data is not fully reliable, as, e.g.,  the
United Nations Agency  for international Telecommunication Union (ITU) self-admits it:

{\it ''The STAT Unit verifies and harmonizes data, carries out research, and collects missing values
from government web sites and operators' annual reports, particularly for countries that do not
reply to the questionnaire. Market research data are also used to cross-check and complement
missing values''} (ITU ICT Statistics Database).

\item
On the other hand, IG is an idea, proposed from a theoretical point of view that is
suitable for the interests of corporations that disregard the role of states at the time of
defining policies, especially about their regulation.
in this context, data analysis followed by the construction of adequate physical
models which are able to describe different future scenarios that can be further
discussed, appear to be if not extremely relevant, very much rewarding.
\end{itemize}

The present  starting point is the survey of the ongoing policies and proposals, such as those
that can be considered to be the outcomes of the WSIS, for instance, the internet
Governance Forum (IGF); see Sect. 2.
One main aim is to give {\it a priori} probable future scenarios, to suggest  more
adequate instruments to avoid DD and to design policies for IG. in order to do so we
have used two much worked upon models taken from statistical physics to investigate each
case respectively in Sect. 3. For DD, we use  Huang version of the Ising model
(Huang 1967); for IG, a multiagent ( a so called prey-predator approach) system (Lotka 1925; Volterra 1931) is considered.

\section{The Information Society }
Beyond the concept of Information Society posed when {\it
"... the OECD acknowledged that the economy of tomorrow will be, to a great extent,
Ôinformation economyÕ and the society will become Ôinformation societyÕ which means that
information will account for a significant part of added value of most goods and services
performed and information intensive actions will become distinctive of the households and the
citizens" } (WSIS 1998),
  leaving aside 
the discussion about whether it is or should be
named  ÔKnowledge SocietyÕ, rather than  Information SocietyÕ,  the fact remains that today this
concept involves important social, legal and economic issues, many of them unexpected in the nineties. The more so nowadays due to the plethora of economic means, when a newly evolving eIS expands.

\subsection{About policies}
There is a lack of a consensus on the definition of public policy (Birkland  2001). The meaning here given to the word ÔpolicyÕ is a course of action or inaction chosen
by public authorities to solve, in most cases, an interrelated set of problems, as summarized by $http://psychology.wikia.com/wiki/Government_{-}policy_{-}making$.here below only considering  such problems
derived from the complex nature of the Information Society. One main question  arises:  can
such problems  be solved by means of policies for ICT or  rather should
  policies be implemented  for the Information Society {\it per se}?
Considering that Òthe Information Society ÔwillÕ affect most aspects of our livesÓ,
European policies, e.g.,  range from the regulation of entire industrial sectors to the
protection of each individual's privacy  (European Commission 2010)\footnote{$http://ec.europa.eu/information_{-}society/activities/broadband/policy/index_{-}en.htm$}. In Latin America,
current programs for IS are likely {\it Òto promote public policies for the advancement of
development-oriented IS, by aligning policies on the use of ICT for development; to
promote transparent and participatory interaction Ó }(ECLAC)\footnote{$http://www.eclac.org/socinfo/acerca/programa/default.asp?idioma=iN$}. 

Whence, for different realities, the
same need of policies occurs. Moreover, the complexity of the governance appears since such policies should be both adequate to each
society, i.e. locally, beside being globally workable.

\subsection{ The  Digital Divide (DD)}
However one should not assume the world to be so uniformed. Among the problems regarding the IS, and probably the most conflICTive of them, is
the problem that makes the world  be divided into people who do have and people
who do not have access to modern information technologies. This problem is called
the Digital Divide (DD). As the examples below do show, the DD has no univocal
meaning.

In Europe (EU), the problem is considered in terms of "Broadband Gap Policy", which is
concerned with the geographical aspects of the digital divide among EU regions. 

In
the USA, the DD seems to be analyzed in terms of individual options: {\it 
Òa study published by the Pew internet \& American Life Project has found that there is a
growing digital divide across America. Whilst a reasonable number of Americans are embracing
new technology and Web 2.0, a disturbing number are either not getting the message, or are
choosing not to participate.Ó }(Riley 2007)

For Latin America (LA), the relation digital divide/social coherence is considered relevant
not only by social scientists, but also by international organizations such as CEPAL
(Hopenhayn 2008). However, corporations restrictedly analyze the problem in terms of
telecommunications infrastructure (OECD). 

In Africa where for most people
even making a telephone call is still a remote possibility, cellular phones and internet
telephony are considered as if they were Ô{\it taking on the Digital Divide}.

In short, given the reality of todayÕs competitive socio-economic scenario, societies
appear to be divided between those who are"Ôin" (included) and "out of" (either
excluded or not included) the e-IS. But, beyond the ideas of those that analyze the world
as if it were all alike Europe (Derrida 1987; Foucault 1984), the DD involves the gap
between the educated and uneducated, between economic classes, and the more and
less industrially developed nations. The Dependency Theory has demonstrated that
the cause of such a dual society is the lack of endogenous growing capability
(Cardoso and Faletto 2001). From another theoretical position, it has also been proved
that whereas the developed societies have endogenous growing capability, the
underdeveloped ones lack it (Romer 1990). So, policies are required to bridge the
gap, 
beside giving rules for the evolving "game".

\subsection{The internet Governance Forum}
From a theoretical viewpoint, it is unconceivable that the internet, being a global
network, should be submitted $only$ to the national state regulation of each connected
country. Neither should it be submitted to the national state regulation of $one $ given
country. in fact, internet would be an ideal example of an institution that can only be
ruled by international law; but it is not. 

Neither scientific nor political significant
efforts are being made in this direction.
On  one hand, the WSIS is committed to governance, - a term that corresponds to the
so-called post-modern form of economic and political organizations.  Recall that "Governance" has,
at least, six different meanings: the minimal State, corporate governance, new public
management, good governance, social-cybernetic systems and self-organized
networks (Rhodes 2007). Most of these viewpoints focus on legitimating projects of neo-liberal
inspiration (de Senarclens 1998).
{\it ÒGovernance can be seen as the exercise of economic, political and administrative authority to
manage a country's affairs at all levels. it comprises the mechanisms, processes and institutions
through which citizens and groups articulate their interests, exercise their legal rights, meet their
obligations and mediate their differencesÓ } (UNDP).

On the other hand, some authors maintain that the cyberspace shows a somehow
feudal character that emerges from the hierarchical privatization of its government
associated with the granting of internet domains (Yen 2002; Elkin-Koren and Salzberger 2004).
That is why the internet's government, like that of a feudal society, is highly
fragmented (Yen 2002).

In the Tunis phase of the WSIS, in November 2005, governments asked the UN
Secretary-General to convene a Forum, with the mandate to discuss the main public
policy issues related to internet Governance in order to foster the internet's
sustainability, robustness, security, stability and development. (WSIS-
05/TUNIS/DOC/6(Rev.1)-E). The inaugural Meeting of the internet Governance
Forum (IGF) took place in Athens, in November 2006;  in November 2007,  a
Second Meeting took place in Rio de Janeiro.

The different criteria for listing the participants (see Provisional Lists) in  such meetings
prevent us from making an accurate comparison of both meetings, in terms of "cause" and "effects".  Taking a binary polarity point of view as a first approximation, it can be deduced, however, that
companies, trade associations and non profit organizations fully committed to the
internet were, in both of these meetings, composing the majority among the ÔEntitiesÕ, - a wide  group, while a somehow
Ôprivate sectorÕ category that seemed  or was  supposed to represent the Ôcivil societyÕ was the minority group.
Notice that, such organizations can also be found among ÒInternational OrganizationsÓ,
provided they have any kind of international activity, e.g. ISOC, italy being an example of
this statement.

Leaving aside, therefore in the present considerations, the current discussion on state or non-state regulation (de Souza Santos and Rodriguez Garavito 2005), there is nevertheless no doubt that there must be some kind of
regulation regarding the internet, and that such a task demands 

{\it
Òthe full involvement of governments, the private sector, civil society and international
organizationsÓ } (WSIS, Tunis Agenda, 2005, 2).

and

{\it 
Òin addition, there is a need to consider the following other issues, which are relevant to ICT for
development and which have not received adequate attention: ÉActivities on ICT-related
institutional reform and enhanced capacity on legal and regulatory frameworkÓ }(WSIS, Tunis
Agenda, 23, j).

So be it. interestingly, the Chairman's Summary of the second meeting provides  some good material to analyze
the debate regarding the legal aspects involved. The document shows, among several
appeals to self-regulation and soft law instruments, a consistent demand of state
regulations. An ECLAC document (Newsletter N$^{\circ}$4) informs us, however, that

{\it 
ÒRepresentatives of Brazil introduced a variety of proposals about how to reform the ICANN,
restrICTing its function to that of a coordinating organization. There was an extensive discussion
about the future of the ICANN and its relationship with the USA Department of Commerce,
whereas Brazil recommended the creation of a new international agency, composed by
representatives of the civil society, to rule the accessÓ.}

The Chairman's Summary alludes to this debate, but alas records neither authors nor
proposals.

{\it 
ÒOther points covered the relation of governments to ICANN and whether is was appropriate for
the Government Advisory Committee (GAC) to have only an advisory role as opposed to fuller
powers in terms of international public policy. While one panellist argued that the participation
of governments in the GAC was one of ICANN's most important features, another put forth that
the current model with GAC as part of ICANN was not a stable model.Ó}

\subsection{State Actors}
ÒA vast literature has developed over the last few years that theorizes and empirically studies
novel forms of governing the economy that rely on collaboration among non-state actors (firms,
civil organizations, NGOs, unions, and so on) rather than on top-down state regulation. É From
this viewpoint, the solution lies neither in the state nor in the market, but rather in a third type of
organizational form, i.e. collaborative networks, involving firms and secondary associationsÓ (de
Souza Santos and Rodriguez Garavito 2005).

Let the following quote be emphasized: 
 
{\it 
ÒThe international management of the internet should be multilateral, transparent and
democratic, with the full involvement of governments, the private sector, civil society and
international organizationsÓ} (WSIS, Tunis Agenda, 2005, 29).

This is merely one of several similar paragraphs that can be found in WSIS
documents. Such seemingly horizontal and democratic statements hide the fact that,
apart from state and market, only the elites or members of the middle-class with the
economic and cultural capital shall be stakeholders in the IG (de Ort\'{u}zar {\it et al.}  2007).

\section{ Mathematical modelizations of the Information Society}

A mathematical modelization would help  visualizing if policies could influence
the behaviour   of social agents or not. in order to attain this purpose we
have chosen the Ising model  
and a multiagent, prey-predator-like, so called Lotka-Volterra model (Lotka 1925; Volterra 1931).

\subsection{ Ising model approach to Digital Divide}

The Ising model is one of the pillars of statistical mechanics. When developed by
analogy in  sociophysics one considers that the world is composed of lattice  sites  on 
which are located agents (originally magnetic moments); they ÒinteractÓ with others in
their neighbourhood; each site (agent) can have two values i.e. +1/-1; for simplicity
we will consider that the underlying site network is a square lattice. Here we use it as
a neighbour behaviour model. As in the Ising original version, the present model is
analyzed in terms of two parameters: the temperature of the system and the external
field. The temperature is associated to the degree of interest or relevance
concerning a given situation, in this case to be ÔinÕ or ÔoutÕ of the IS, while the
external field(s) represent(s) the applied policy(ies). For simplicity, here below,  the temperature is supposed to be the same in all  
simulations. 
it is intended to examine  how policies (the external fields) are able to accelerate the arrival
at a desired situation, i.e. in the present case to have more agents ÔinÕ than ÔoutÕ the IS.

The lattice, which represents a given ÔsocietyÕ, has 100 x 100 agents. The initial
situation in each simulation is taken at random, between +1 or Ð1 for the initial agent
values.  The simulations have been made using the method developed by Caiafa and
Proto (2006) here below for three different enforced policies. The evolution in time (arbitrary units/iterations) without an external field (H=0)  is
considered to be a situation without policies.  
An external field, e.g., 
H = 1, in arbitrary units,  corresponds to a weak policy enforcement.  A more adequate or stronger
policy can be also considered, e.g. for an external field H = 2 in arbitrary units. it is possible to follow the
evolution within the ÔinÕ and ÔoutÕ agents   as a function of time in the  three cases. The images and plots of the evolution of
number of agents with value +1/-1 is shown, in Figs. 3-7, indicating that the appliance of adequate
policies  drastically reduces the evolution time required for the society  in order to arrive to a state with  more ÔinÕ than ÕÕoutÕÕ agents. Comparing  the data on the
  figures shows that if an external field is not provided (H=0),  a lower number of
agents is found ÔinÕ;  (ii) it takes around three times longer to reach   the same
evolution  state when H goes from 0 to 2.

Taken as example an interacting agent  society (Fig. 1) in which several agents are in,
and others out, initially distributed at random in the society (Fig. 2), and leaving the system to evolve  without  any imposed  policies, the ÔinÕ and ÔoutÕ agents nucleate  in order to form two well defined
fields. At the end, considered the asymptotic limit, the population of both categories is more or less the same and appears to be
bound to stay still if the simulation time is increased. This is a DD state (Fig. 3). Now, by means of applying certain
policies, the analyzed society acquires more mobility (Fig. 5  and Fig. 7). The agents offer resistance
against the policy, what is shown in the strong transitory with rises and drops (Fig. 6 and Fig. 8). Finally, some of them move slowly from ÔoutÕ into ÔinÕ, but the final result of such
movement, though faster for H=2, is not much different from H=1. 

in practical words, it seems that the particular society under analysis has been provided with the required
induction. in other words, the policies here adopted (H=2) have proved to be more adequate
than the previously more simple one (H=1).

  \section{Results}
 
  \subsection{Some results of actual policies}
 it is of interest to recall whether some virtual features as those found above have had some similarity in the real world. in fact, in 
 2006, in Sao Paulo 54\% of the entrepreneurs had access to internet,
but only 47\% of their firms had at least a personal computer; their access was used
for on line bank and governmental services and e mail in an 83\% (Bede 2003).
On the other hand, the Argentinean firms in GBA not only had the equipment in the
assets -93\%- but also access to internet -90\%-, but made poor use of it -62%
procedures, 49\% on line banking services and 80\% e mail- (UNLP 2005). Why is it
that Brazilian entrepreneurs did such an intense and dynamic use of ICTs, even
ÔoutsourcingÕ the access to internet? The answer might be in the policies the Brazilian
government adopted.
in Brazil the Information Society, e-government being a part of it, is a state policy
(Wilson 2004), not Ôone governmentÕsÕ policy. Consequently, successive
governments have developed a strong program based on early decisions aimed to
discourage physical presence when on line procedures are possible. it seems that
Brazilian policy makers have found the adequate external field Ðpolicy- in order to
accelerate the arrival at a desired situation -agentsÕ dynamic interaction using ICTs,
even by means of someone elseÕs equipment-.

  \subsection{ Internet regulation as a multiagent system.  Lotka-Volterra model}
 
To give a more concrete exemplification of the discussion on state or non-state
regulations, and particularly to enhance the importance of the participation of non-state
actors, it is adequate to appeal to a simple semi-empirical modelization of the
problem at hand. The model introduced by Lotka (1925) and Volterra
(1931) is applied  here below  to take into account a weight  for (or size of ) for agents, going beyond the binary polarization hypothesis studied here above.  Such an extension of the Lotka-Volterra model outside biology and anthropology has been used in related fields to the present study, like in order to model the competition between web sites
(Maurer and Huberman 2003),  in hung scenarios in sociology (Caiafa and Proto 2006).
The set of $N$ differential equations (Maurer and Huberman 2003) of the model is the
following:
  
 \begin{equation}
\frac{df_i}{dt}= \alpha_{i}f_{i}\left( \beta_{i}-f_{i}\right) -{\displaystyle%
\sum\limits_{i\neq j}} \gamma\left( f_{i},f_{j}\right) f_{i}f_{j} 
\mbox{ \ \
\ \ \ \ for \ \ \ \ } i=1,..,N  \label{Model}
\end{equation}
where $\frac{df_i}{dt}$
means the time derivative of $f_i$, and indexes $ i, j$ run from 1 to $N$. The $f_i$
 is the weight of the $i$ agent opinion, at time $t$, such that it is taken from  a normalized $f_i$ distribution. in other words, if some agent has an increase in size, another or a group of others must have a corresponding decrease in size. The
parameters of the model are: $\alpha_i$, the growth rate of agent $i$, and $\beta_i$, the saturation value
of the $ i$-th  agent. in order to introduce the effect of the ÔsizeÕ of the agents,  one can 
redefine the growth rate parameter $\alpha_i$, according to Economo {\it et al.} (2005) as:

 \begin{equation}
 \alpha_{i} = \left(\frac{a}{b_i}\right)^4 
 \label{pressure}
\end{equation}
where $a$ is the selection pressure which is, for simplicity, hereby, taken to be equal for
all the agents living in the same environment, here the whole  Information Society, and $b_i$ is
a parameter which reflects the inverse of an agent competitiveness; for example, in
organization theory,  $b_i$  is associated to the  {\it cost to do something} (Porter 1980). in
the present study, the competitiveness should be understood as the cost imposed to an agent 
ideas/interests in order  to  accept  the  imposed regulations on the Information Society. This
modification of the growth makes it possible to take each agent  'size' into account as
suggested by Economo {\it et al.} (2005).

in our modelization, we have introduced two kinds of agents:
\begin{itemize}
 \item 
 The Õwell-established in the Information Society agentsÕ; let us call them Old
(O); they are e.g. ICANN, i.e. the gatekeeper of the internet, software companies,
internet providers and NGO involved in the development of communications
and the internet. These agents presently lead the {\it de facto} management of the
net. We call them stakeholder.
 \item 
  The agents that are trying to find a seat in the Information Society
Governance. These are Ôcivil society agentsÕ, like NGO, individuals, SME and
the like. Can be also included in this category, several governments that still
have no definite policy about IS
We call them New (N)  or participant.
\end{itemize}

\subsection{Simulation results} \label{sec:Simulationresults}

For the present work, in order to illustrate the analysis, we consider only ten agents,
keeping $a$ =1 equal for all agents which means   that all agents, living
in the Information Society, are equally supporting the selection pressure. This means that, ideally, all the agents have the same rights as regards to  the
policies for the sustainability of the Information Society. Whence only the agent
competitiveness has to be varied in order  to look at the evolution and to determine the long term weight, i.e.
importance of its opinion $f_i$, of the $i$-agent. Also for simplicity we keep $\beta = 1$. The $\gamma_{i,j}$ values are fixed and all equal either to +1 or -1. The O agents are supposed to be in competition among
their community, through their $\gamma_{i,j}$ =-1), taken to be the same in all simulations. in
Figs.  9 to   12,  the evolution of $f_i$ is shown for the exemplary case where there are 40\% O agents, each having a different $\beta_i$ = 0.10, 0.11, 0.12 to 0.13 in presence of 60\% N agents, each
having a different $b$ ranging from 0.41 to 0.46, thus differing by 0.01 steps (0.4, 0.41, 0.42, 0.45,
0.46). The initial condition on $f_i$ for both, the O- and N- agents is equal to 0.1. 

The
simulations show that when N do not cooperate among themselves (Fig. 9), their
weights  remain always below all the Old's weight. Therefore their opinions remain  
irrelevant or do not count in face of the O's opinions. in such a scenario, Ògovernments,
the private sector, civil society and international organizationsÓ, that are not well
established actors in the Information Society yet, in other words, mere
PARTiCiPANTS, do not have any chance  that their demands be attended, as regards to the
Information Society's regulation. Nevertheless,  it was   also found that, when
N agents cooperate among themselves, the weight of the opinions of O and N get
closer and closer to each other: there is even a possibility for the opinions of N to win (Fig. 12)
when all of them are in a cooperation scheme.

In these figures, it is demonstrated that cooperation allows New agents to become more and more powerful.

\subsection{Interpreting policies in the light of the model} \label{sec:interpret}

Coming back to the internet Governance context, in the Chairman's Summary we
read:
\begin{itemize} \item {\it
ÒThere was a clear convergence of views that governments had an important role to play in
creating a solid regulatory framework and making sure that the rule of law was well established
and respectedÓ.}
\end{itemize}

It can be asked:  convergence of whose views? in view of the presented
figures, it may be wondered also whether it is a convergence of views of individuals, of
organizations or of contributors. One can also question in whose benefit has such a
demand been posed?

As the simulation results show, there is a chance to achieve a scenario where, by
means of cooperation among N agents, their demands are attended. There are many
weak agents among the N, but there is also China, Brazil and some relevant
independent NGO. From such a further developed scenario might emerge the rules
for a widely comprehensive and satisfactory government of the internet. it might also
set the basis for the legal and political frames of the Information Society.

\section{Conclusion} \label{sec:concl}

Numerical simulations with the Ising model let us see that the appliance of adequate
policies reduces drastically the time required for the society under analysis, to arrive
at, almost, all ÔinÕ agents. if external field is not provided, a lower number of agents
shall be found ÔinÕ, and it takes around three times greater evolution range. Thus we
have learned that, in many situations Ðcountries, regions, social groups-, defined
policies should be implemented in order to encourage people to move ÔinÕ the
Information Society.
Through the multiagent system, we arrive to the conclusion that to attain an active
role in the Information Society, and therefore participate in policies decisions, the N
agents should cooperate among themselves (see Figs. 10-12)
  We have even some
indication of the order of magnitude of the number of necessarily cooperating N
agents in order to overthrow the O opinion/attitude/ and how long it takes (see Figs. 10-12).
in both cases simulation results lead to solutions which are clearly equivalent to the
consequences that some social scientists have forecasted, in terms of theoretical
explanation of phenomena that are actually comparable to those that are taking place
within the Information Society.
in summary, the social modelization of the Information Society as a complex system
provides insights about how the Digital Divide can be reduced and how the huge
majority of ÔweakÕ members of the IS would influence the outcomes of the IG and, in
so doing, allow the internet Governance to Òbe multilateral, transparent and
democraticÓ.

   \vskip 0.5cm
  
\newpage
Bed\^e, Marco Aurelio. 2003,  A informatiza\c{c}ao nas MPEs paulistas, (S\c{a}o Paulo: SEBRAE ) .  \\

Birkland,  Thomas A.. 2001, An introduction to the Policy Process: Theories, Concepts, And Models of Public Policy Making (Armonk,  NY: M.E. Sharpe), p.21.\\

  Caiafa, Cesar F. and Araceli N. Proto, 2006,  Dynamical emergence of contrarians in a 2-D Lotka ÐVolterra lattice, {\it international Journal of Modern Physics C},  17: 385-394\\

 Cardoso,  Fernando Henrique and  Enzo Faletto. 2001, Dependencia y desarrollo en Am\'erica Latina, Siglo XXi Editores Argentina/Mexico (1$^{a}$ed.1969).\\

CEPAL, Programa Sociedad de la Informaci\'on $http://www.cepal.org/SocInfo/ $\\

CEPAL ,  Newsletter N$¬^{\circ}$4 Sociedad de la Informaci\'on. Temas destacados del
Segundo Foro de Gobernanza de internet, Enero-Febrero, 2008, 8.\\

de Ort\'{u}zar, Maria G., Noemi Olivera   and Araceli Proto. 2007,  Justice and Law in/for the
Information Society, Actas de CollECTeR Iberoam\'erica, Argentina, pp. 297-304.\\

Derrida, Jacques. 1987,  La escritura y la diferencia (Barcelona: Anthropos). \\

de Senarclens, Pierre. 1998, Mondialisation, souverainet\'e et th\'eories des relations
internationales (Paris: Armand Colin).
   \\

de Sousa Santos, Boaventura  and C\'{e}sar A. Rodr\' iguez-Garavito. 2005, Law, Politics and the
Subaltern in Counter-Hegemonic Globalization, in Law and Globalization from
Below: Toward a Cosmopolitan Legality', Cambridge Studies on Law and Society, (Cambridge,MA: Cambridge University Press). \\

  ECLAC Programme for The Information Society \\

Economo, Evan P., Andrew J. Kerkhoff,   and Brian  J. Enquist. 2005,  Allometric growth, life-history invariants and population energetics,  {\it  Ecology Letter} 8: 353-360. \\

Elkin-Koren, Niva and  Eli M.  Salzberger, 2004, Law and Economics and Cyberspace. The Effects of Cyberspace in the Economic Analysis of Law (Cheltenham, UK: 
  Edward Elgar).
 \\

European Commission. 2007, Governance in the EU. A White Paper. \\

European Commission. 2008,   2010-Broadband Gap. Broadband gap policy. \\
  
 European Commission.  2010. Information Society and Media. Strategy for an
innovative and inclusive European Information Society
 \\
 
Foucault, Michel. 1984, What is Enlightenment? ("Qu'est-ce que les Lumi\`{e}res ?"), in Rabinow, Paul, Ed., The Foucault Reader, (New York: Pantheon Books), pp. 32-50. 
   \\

Hopenhayn, Martin. 2008, Brecha Digital y Cohesi\'on Social en Am\'erica Latina Una
ecuaci\'on que no cierra, Seminario sobre los Retos de la Brecha Digital en Am\'erica
Latina,  Programa EUROsociAL, Santiago de Chile. \\

Huang, Kerson. 1967, Statistical Mechanics (New York: J. Wiley).\\

IGF internet Governance Forum. 2007, Chairman's Summary Second Meeting, Rio
de Janeiro, see
 $http://www.intgovforum.orgcms/secondmeeting$ (Chairman Summary.
FiNAL.16.11.2007.pdf) \\

IGF. 2006, Provisional List of Participants, $http://info.intgovforum.org/PL.php$ \\

IGF. 2007, Provisional List of Participants, $http://info.intgovforum.org/PLP_{-}2IGF.php$\\

ITU ICT Statistics Database, $http://www.itu.int/ITU-D/ICTeye/Indicators/Indicators.aspx$ \\







Lotka, Alfred J.. 1925, Elements of Physical Biology,  (Baltimore: Williams \& Wilkins). \\

Maurer, Sebastian  M. and  Bernardo A. Huberman. 2003, A Competitive Dynamics of Web Sites,  {\it Journal of Economic Dynamics and Control} 27(11-12): 2195-2206. \\

Mutume, Gumisai. 2003, Africa takes on the digital divide; new information technologies
change the lives of those in reach. {\it Africa Renewal} 17(3):
 \\

  OECD. Latin American Economic Outlook, see

$http://www.oecd.org/department/0,3355,en_{-}2649_{-}33973_{-}1_{-}1_{-}1_{-}1_{-}1,00.html
  $ \\


Porter,  Michael. 1980, Competitive Strategy (New York:Free Press).\\

Rhodes, Rod A.W.. 2007,  Understanding Governance: Ten Years On,  {\it Organization
Studies} 28(8): 1243-1264. \\

Riley, Duncan. 2007,  America: The Growing Digital Divide, Tech Crunch, 

$http://www.techcrunch.com/2007/05/06/america-the-growing-digital-divide/$ \\

 Romer,  Paul M.. 1990, Endogenous Technological Change, {\it Journal of Political Economy} 98(5, pt.2): S71-S102 \\

UNDP. 1997, Governance for sustainable human development. A UNDP policy 
document, New York, 1.2. \\

UNLP. 2005, Observatorio de Tecnologias de informaci\'on: Estudios realizados con
Pymes (oct./dic. 2004 y oct./dic. 2005) \\

Volterra, Vito. 1931, Le\c{c}ons sur la Th\'eorie Math\' ematique de la
Lutte pour la Vie  (Paris: Gauthier-Villars). \\

 Wilson, Ernest James III. 2004, Strategic Restructuring in Brazil, in: The information
Revolution and Developing Countries, (Cambridge, MA: MiT Press),  pp. 119-172.\\
  

 World Summit on the Information Society $http://www.ITU.int/WSIS/index.html$\\

Yen, Alfred Chueh-Chin. 2002, Western Frontier or Feudal Society? Metaphors and Perceptions of Cyberspace, {\it  Berkeley Technology Law Journal} 17: 1207-1263.  
   
  


 \newpage
\begin{figure}
\centering
\includegraphics[height=8cm,width=8cm]{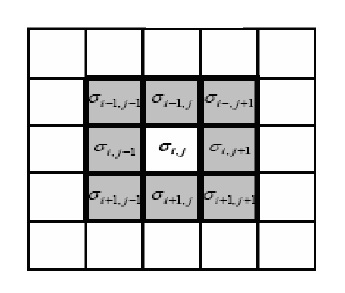}
\caption{\label{1}  Neighboring structure of  an  agent  located at site $i,j$ : each one has
eight  neighbors with whom to
interact. Each agent is in a state $\sigma$ depending on its location}  
\end{figure}

\begin{figure}
\centering
\includegraphics[height=8cm,width=8cm]{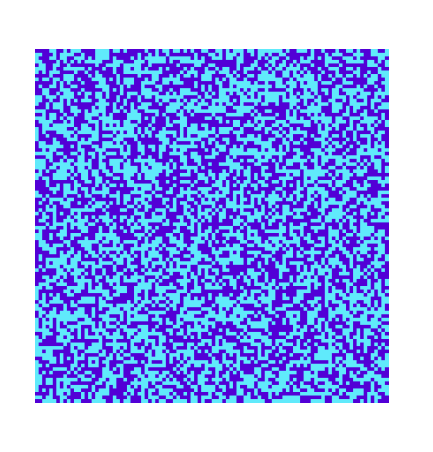}
\caption{\label{2}  Bidimensional plot showing
initial situation of each agent.
Blue dots are the ÔinÕ agents
(+1) and cyan the ÔoutÕ(-1)
ones, as in the following
draws.}  
\end{figure} 

\begin{figure}
\centering
\includegraphics[height=5cm,width=5cm]{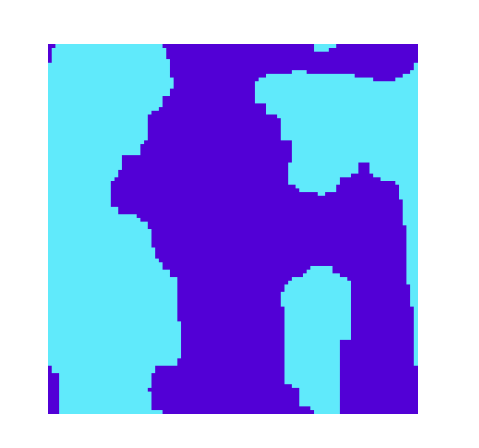}
\caption{\label{3}    Bidimensional plot showing
the ÔÕstableÕÕ situation reached by each
agent  without an external field
(without policies)}  
\end{figure}

\begin{figure}
\centering
\includegraphics[height=5.6cm,width=5.6cm]{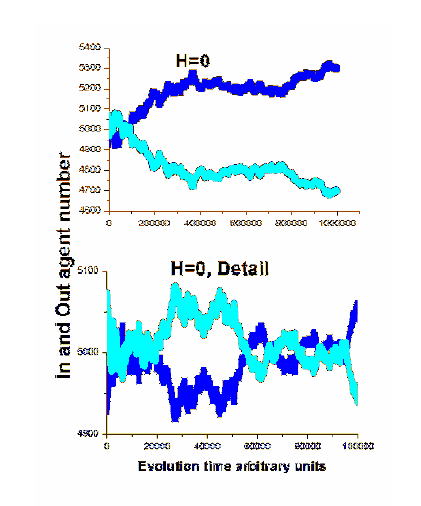}
\caption{\label{4}   The number of ÔinÕ and
ÔoutÕ agents, plotted versus
time/iteration  without an external field
(without policies)}  
\end{figure}

\begin{figure}
\centering
\includegraphics[height=5cm,width=5cm]{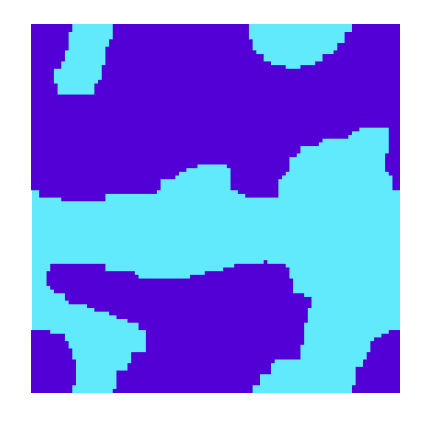} 
\caption{\label{5}    Bidimensional plot showing
the ÔÕstableÕÕ situation reached by each
agent  with an external field (H =
1, arbitrary units), i.e. some  
policy, is applied}  
\end{figure}

\begin{figure}
\centering
\includegraphics[height=5.6cm,width=5.6cm]{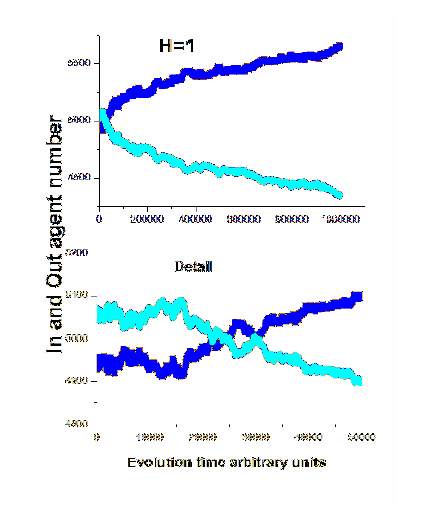}
\caption{\label{6}   The number of ÔinÕ and
ÔoutÕ agents, plotted versus
time/iteration with an external field (H =
1, arbitrary units), i.e. some  
policy, is applied }  
\end{figure}

\begin{figure}
\centering
\includegraphics[height=5cm,width=5cm]{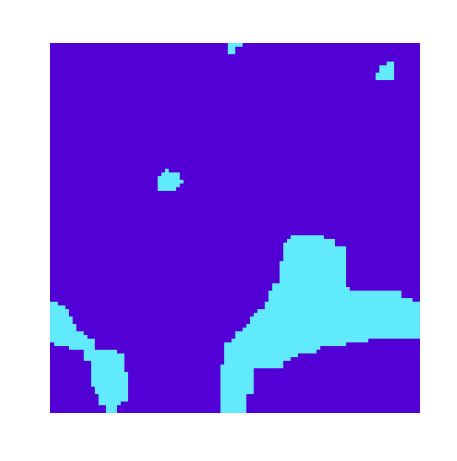}
\caption{\label{7}    Bidimensional plot showing
the ÔÕstableÕÕ situation reached by each
agent when an external
field (H = 2, arbitrary
units), strong policy, is
applied }  
\end{figure}

\begin{figure}
\centering
\includegraphics[height=5.6cm,width=5.6cm]{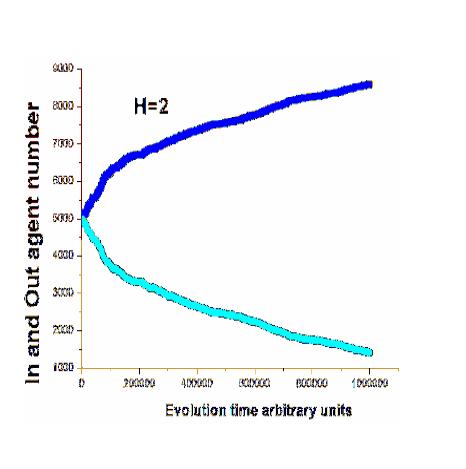}
\caption{\label{8}   The number of ÔinÕ and
ÔoutÕ agents, plotted versus
time/iteration when an external
field (H = 2, arbitrary
units), strong policy, is
applied }  
\end{figure} 

\begin{figure}
\centering
\includegraphics[height=8cm,width=8cm]{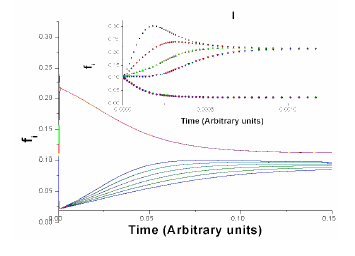}
\caption{\label{9}  New agents do not cooperate and Old agents compete among
themselves. Blue line Old agents (superimposed behavior)  I:
Transitory regime. Violet line: New agents (superimposed
behavior). Black, red, green and blue Old agents. As before,
some curves are superimposed.}  
\end{figure} 

\begin{figure}
\centering
\includegraphics[height=8cm,width=8cm]{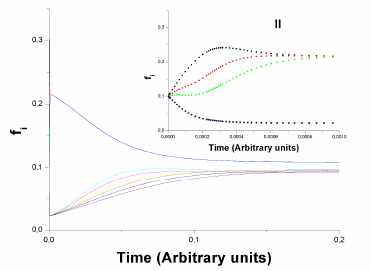}
\caption{\label{10}  Fifty per cent of the New agents cooperate and Old agents compete
among themselves.  II: Transitory regime, colors as in Fig. 9. }  
\end{figure}

\begin{figure}
\centering
\includegraphics[height=8cm,width=8cm]{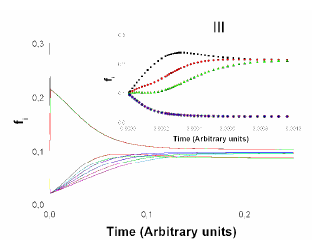}
\caption{\label{11}  Sixty six per cent of the New agents cooperate and Old agents
compete among themselves. III: Transitory regime, colors as
in Fig. 9.}  
\end{figure} 

\begin{figure}
\centering
\includegraphics[height=8cm,width=8cm]{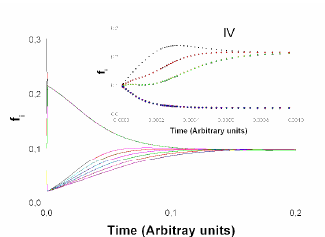}
\caption{\label{12}  One hundred per cent of the New agents cooperate and Old
agents compete among themselves. IV: Transitory regime,
colors as in Fig. 9.}  
\end{figure}

\end{document}